# CERT STRATEGY TO DEAL WITH PHISHING ATTACKS

Shahrzad Sedaghat

Faculty of Engineering Department, Jahrom University, Jahrom, Iran

shsedaghat@jahrom.ac.ir

*ABSTRACT*

*Every day, internet thieves employ new ways to obtain personal identity people and get access to their personal information. Phishing is a somehow complex method that has recently been considered by internet thieves. First, the present study aims to explain phishing, and why an organization should deal with it and its challenges of providing. In addition, different kinds of this attack and classification of security approaches for organizational and lay users are addressed in this article. Finally, the CERT strategy – which relies on three principles of informing, supporting and helping- is presented to deal with phishing and studying some anti-phishing.*

## 1- INTRODUCTION

Phishing is a kind of internet fraud which is defined as an attempt to steal people's sensitive information such as username, password, and credit card information by foisting itself as a trusted website. This attack is a sample of social engineering technique, in which the key to success is the power to gain individuals' trust. Further, it is obvious that attackers welcome anything that can help them to be appeared as legal and legitimate. These attacks often begin with an email trying to seem legal to the victim and direct him to a fake website or get him to send his information. By sending an email to numerous victims, phishers hope to trap a few of them and their success rate increases statistically [1].

Since a wide range of users is the primary purpose of a phisher, and many of these victims are not informed enough to recognize and deal with such attacks, informing organizations and internet users can decrease the success rate of many of these attacks. On one hand, phishing aims to obtain and misuse personal information, and the disclosure issue of obtaining such information about the people in each country affords a lot of money for the government. However, given the increasing ignorance of internet users regarding internet fraud and the need for academic and scientific specialized centers to support, the CERT specialized center responsible for informing, helping, supporting vulnerabilities and cyber security incidents, was established to meet specialized needs of society, improve technical engineering knowledge regarding the security of computer systems, transform the results to society, create innovative technologies to meet the contemporary needs and problems, and provide opportunities for a better future in the field of cyber security. As a centralized organization and with a combination of part-time and full-time employees, this center tries to investigate the security issues and problems to be ready for a quick and proper reaction to problems and challenges of dealing with internet attacks such as phishing.

## 2- IMPORTANCE OF DEALING WITH PHISHING BY GOVERNMENTS AND ORGANIZATIONS

Governments, military centers, and organizations keep a massive amount of confidential information of their employees and clients, some of which has reduced the security to a product level and by using a simple software product, they can secure the information. Protecting confidential information is a commercial and an ethical and legal need in many cases. The existence of a security issue for the formation of phishing attacks can influence an organization in the following ways:

- Reduced income and increased costs

- Loss of reliability and reputation of an organization

- Disclosure of information to competitors and misusing them

- Disruption of current processes and negative side effects on organization and government activities

Phishing activities can reveal personal information about the clients of an organization. Therefore, it is not pleasant for any organization to lose the trust of its clients and investors by involving in such an issue. In addition, the disclosure of personal information of a client due to the negligence of the organization can result in an individual complaint of the respective organization and even prosecution and bankruptcy.

## 3- CHALLENGES AHEAD IN PROVIDING SECURITY AGAINST PHISHING

- The possibility of creating complex phishing website and illegally selling them in online markets

- Availability of Key-logger software that can secretly record and collect activities related to computer username and password

- Providing the phishers with sensitive information of their accounts by the clients for a false reward for participating in fake surveys,

- Another reason of increasing phishing is a very high return on investment. Phishing has an incredible return on investment.

## 4- WHAT ARE THE STEPS INVOLVED IN SECURING AGAINST PHISHING ATTACK?

The following four trees show the multiple methods of phishing to achieve success. Moreover, some countermeasures have been considered in these trees by which the victim can prevent from attacking progress. The first tree (Figure 1) shows the steps of phishing that are common in all of its formation methods. Open relay mail servers that allow spammers to hide their identity, and throwaway accounts that make it possible to send an email via temporary accounts, and regular email accounts allowing the users to receive an email in Email client software, provide excellent context to start a phishing attack.

Other trees (Figures 2-4) investigate three general methods of a successful phishing, including Trojan, deceit, and spywares [3].

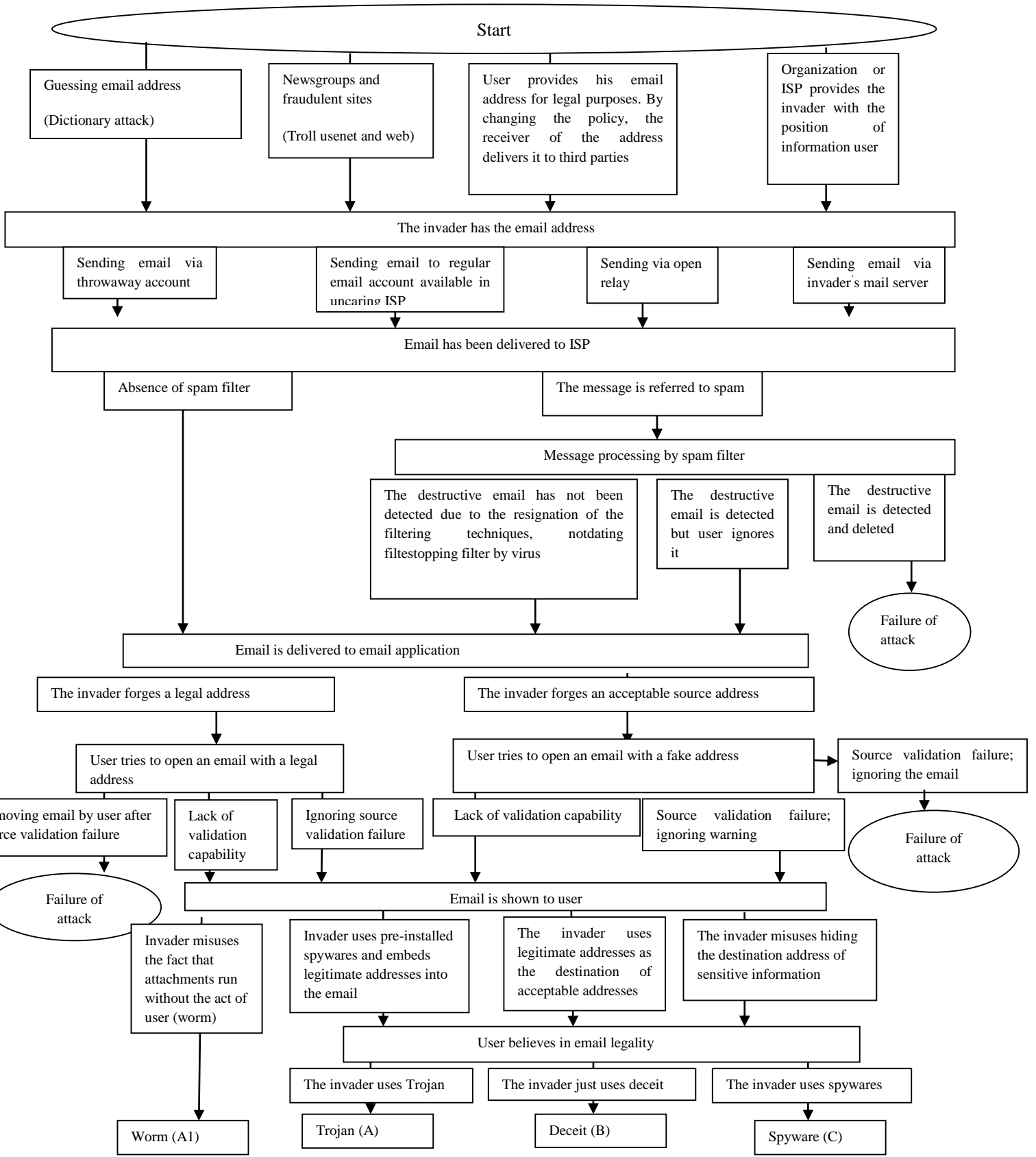

Figure 1- Common sections of phishing attack

Some destructive software may have been attached to the email sent to the victim, that do not behave as he expects and damages his system as soon as he opens them. These destructive codes also spread via pop-up windows. The host-based anti-viruses, anti-spams, and IDSs which register all suspicious movements of the target machine by performing a little service named Agent, and report it to the server via SNMP, have an important role in stopping attacks in this method.

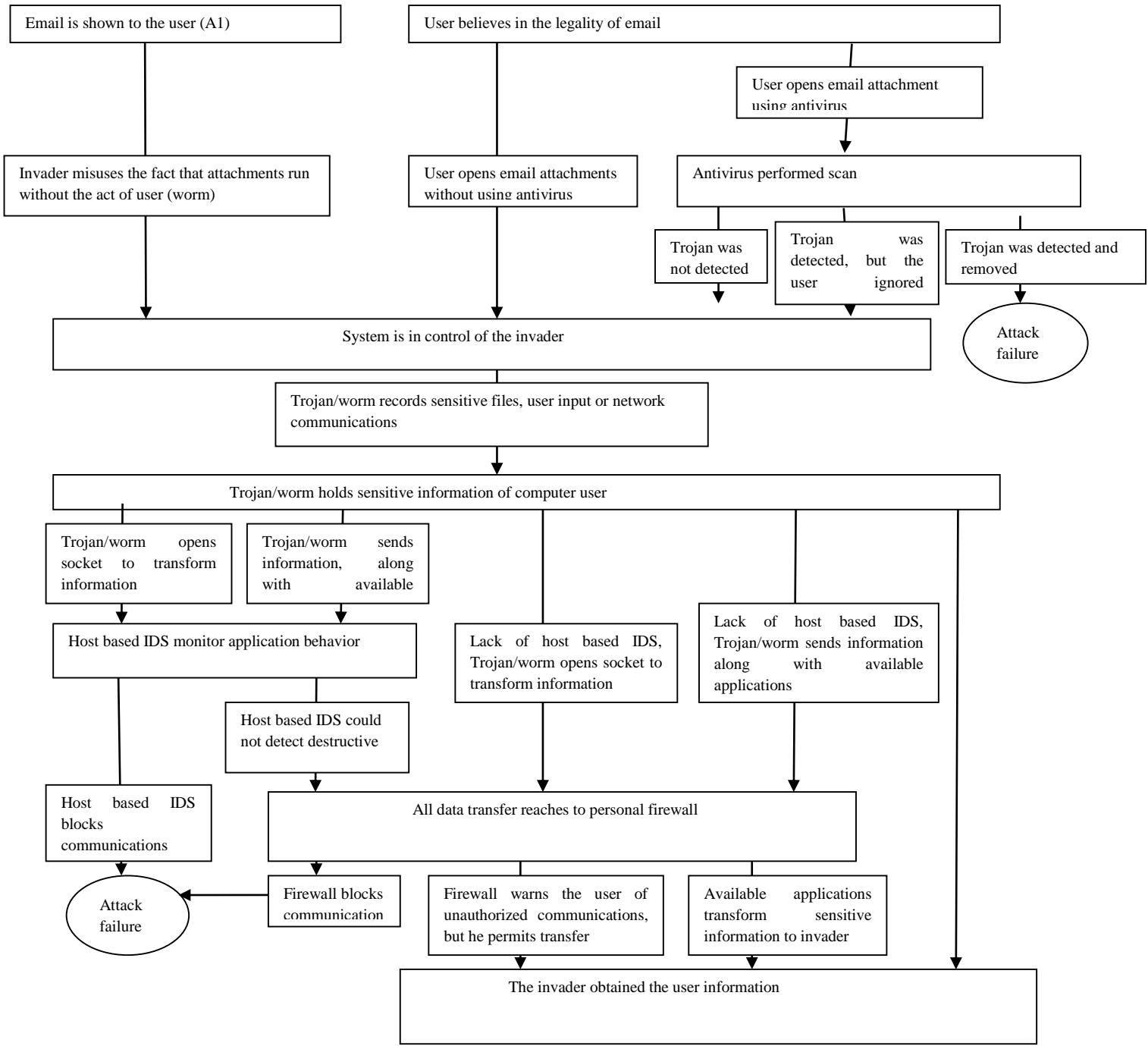

Figure 2- The use of Worm and Trojan in phishing

Relying on deceit and without misusing any additional software, the attacker continues the attack in this method. If the victim trusts the browsers' warnings of an invalid certificate of the website, and the invading website uses an SSL, some protections are taken place.

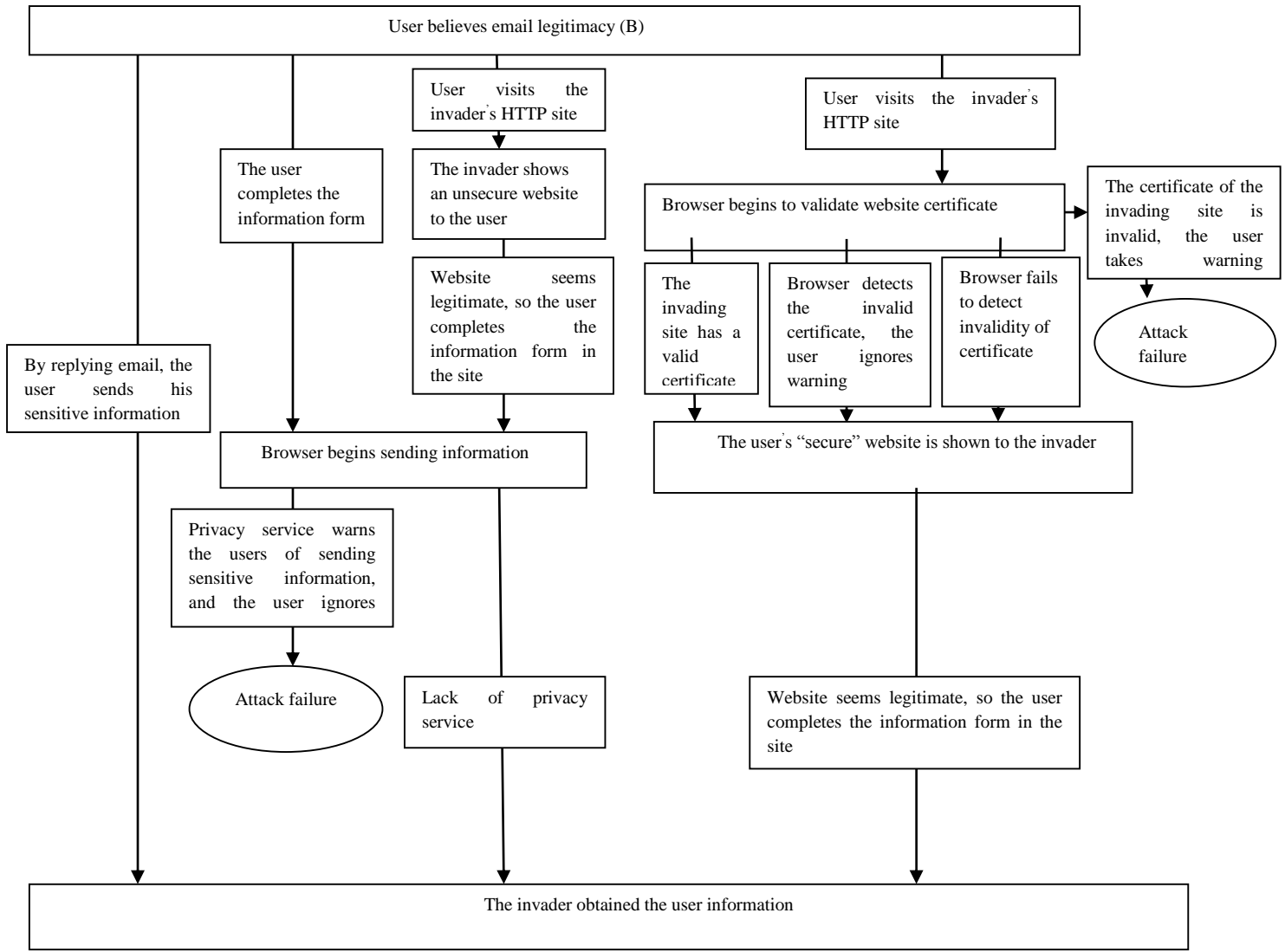

Figure 3- The use of deceit in phishing

In this method, the invader imports a spyware (in the form of a destructive code) to the victim's system. When the victim enters the legal web address intended by the invader, this software activates and directs the victim to the fake website of the invader, or hears the user's communication with a legal website, and provides the invader with the personal information of the victim by any means. Spywares are designed so that they do not perform any demolition or any other action to prevent the victim from suspecting the hearing. Host-based spyware detection specialized software, commercial anti viruses, personal firewalls, and intrusion detection systems can often stop attacks in this type of phishing.

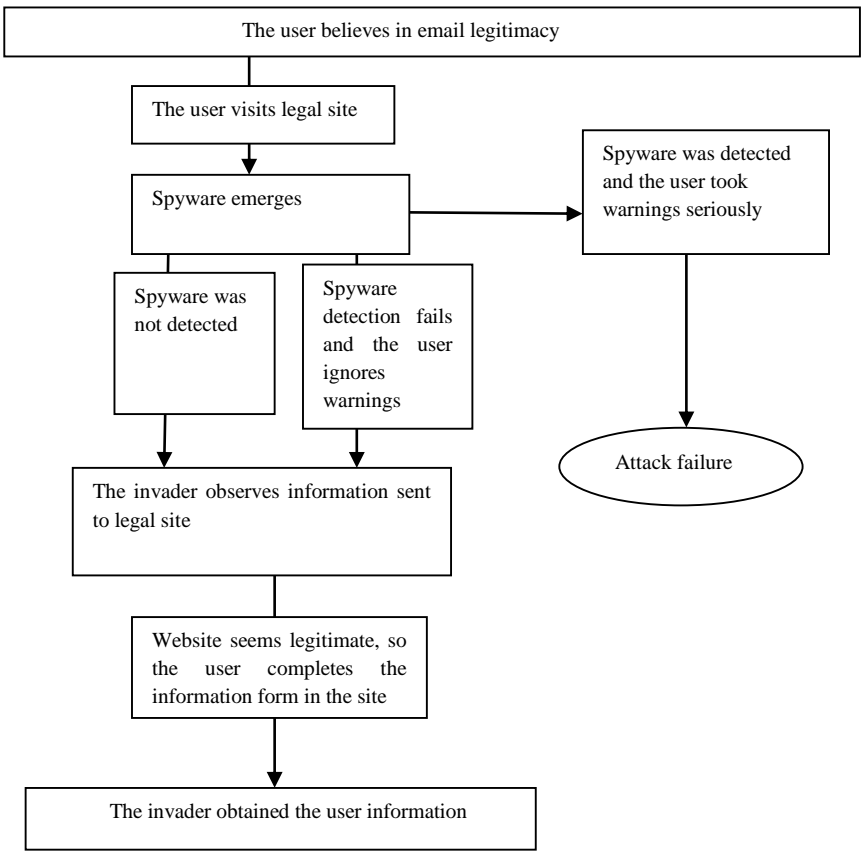

Figure 4- The use of spyware in phishing

## 5. SECURITY APPROACHES TO DEAL WITH PHISHING

Approaches to preventing phishing are generally classified into two groups

### 5-1 Organizational approaches

The primary approach of any organization to deal with any kind of internet attack is to shift the policy and create a relationship between new policies and clients. One of these policies is implementing a method to write the sent emails so that they are not confused with fake emails. In addition, providing users with a solution to validate emails can complement the mentioned policies. Moreover, a victim's information is usually stolen when logging into a website, hence, using a stronger authentication algorithm that does not need entering the sensitive information of the users into organization's website, makes it more difficult for phishers to work. Organizations can provide an additional security layer for their total system. We will explain the key approaches to deal with phishing in organizational environments as follows:

Links that direct the recipient user to a website should not be attached to emails. Although attaching these links provides a simple experience for users, the best secure alternative is culture building the practice of typing and copying the URL into the browser. Organizations should avoid embedding information forms in emails, and never request users for information via such forms. Using such forms makes it difficult for users to recognize the legitimacy of emails. Of course, this approach should be accompanied by educating clients whose emails receiving from the company do not contain any information request form.

Using the Sender Policy Framework (SPF), which is a mechanism to validate the sender of the email in terms of source domain, creates a strong security level for any receiver. This capability is free to use and the steps are such that the server manager should set settings for the organization's mail server in services such as OpenSPF.org. Then, he receives a unique TXT record for his mail server depending on the applied settings and publishes it in the DNS server used by the mail server. Now, the mail server validates all email sender addresses according to the definitions applied to the SPS service and TXT record.

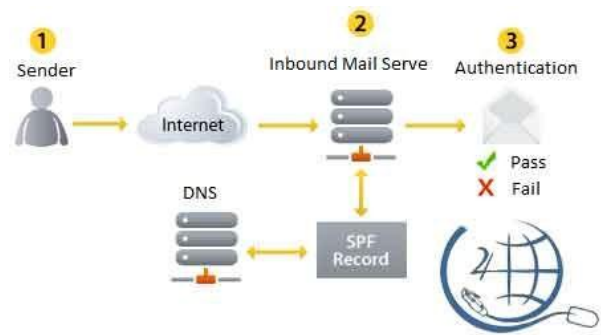

Figure 5- SPF Mechanism

An organization should be able to use the name and photo of the client in writing emails sent for special contracts. Also, it is important to inform clients that every email with a specific content has several contents for validation. In such conditions, the users are able to recognize the limitation of email without installing any additional software or hardware, but the organization is obliged to collect photos of its clients. But, if obtaining the users' photos is impossible or very difficult for an organization, the sequential counting of emails is the best way to create a barrier for phishing. Below is a sample of sequential counting:

- ✓ Date of the current email: Jan. 16, 2017

- ✓ Serial number of the current email: JJH0017

- ✓ Last email we sent to you: JJH0016 on Dec. 10, 2016

- ✓ Serial number of the next email we will send to you: JJH0018

It is worth mentioning that the above approach requires adding serial numbers and an additional security layer to the data base of the company. However, it is still less possible for an organization to prevent phishing, and applying a strong mechanism of authentication can strengthen this possibility. Users can either employ a token creating a disposable password for each authentication to make phishing almost impossible to happen- since they are not aware of the authentication information and cannot accidentally reveal them, or they can choose another method of authentication. Some of the banks are using a technology similar to that of smart cards of GEMP and MAST, which require a connection to another hardware to be detected. In communicating with any organization, the user should carry a token unique to that organization and pay a separate fee for each token. Therefore, using simulator software of token or mobile phone compatible PDAs is recommended to address these approach issues. The application procedure of PDAs is that they create a Pin to be used on the website upon the request of the user. This approach also requires an organizational training plan for a proper configuration of software by clients. Tokens cannot, by themselves, prevent revealing other identification information such as father name, etc. Internet monitoring of fraudulent and phishing websites by computer security organizations, by means of the development of something like white lists and with brand and key content confirmation test and publishing it among internet users can decrease the presence of fake websites over the internet. The monitoring services' providers implement an agent-based solution for an ongoing monitoring of web content and active search of all cases such as logo, brand, or key content, for their clients. Anti-virus scanning at a company gateway of HTTP and SMTP related data traffic can reasonably reduce Trojan, Backdoor programs, and Spywares entrance. This is an easy and quick approach for organizations so that, instead of updating all desktop scanners, they can just update the gateway scanner; however, these two scanners should come together as part of the defense layer strategy against destructive codes since some of encrypted traffics must be scanned at the desktop and mobile phone users are not always protected by gateway scanner while away from their office. With the possibility of the existing phishing websites, organizations should block their access to the internal network by filtering the content and configuring rules blocking destructive URLs in the gateway or using anti-spams. Anti-spams delete or isolate spams in one of the methods of refinement marking of email subject although spammers always change their spamming techniques and no solution is absolutely responsive.

## 5-2 Approaches for lay users

Being suspicious and acting conservatively is regarded as the users' main approach to deal with phishing when entering sensitive information over the internet for any reason (for example, typing the address of any website into the browser and not using links) [4]. If a phishing is powerful enough that users cannot visually recognize the legitimacy of a website or email, there are various software that can help them in this regard. But, it should be noted that this kind of software is not error free, and they may act as a kind of internet attack. Therefore, introducing and instructing buying, installing, maintaining and the way these products work, can ensure a right choice by users. Using anti-spam on desktop and gateway do not allow an attack on spam review by the user. Gateway is for the users of the internet (ISP) and email (MSP) provider companies. There are various methods to inject anti-spam filters into the email processing cycle. One of the three-layer methods involve the following steps:

- ✓ Row 1: installing filter on ISP and MSP for all email client and users

- ✓ Row 2: installing filter for all users on the local LAN network

- ✓ Row3: installing filter on the personal desktop by means of installing security applications

The above-mentioned approach requires the cooperation of ISP and MSP. However, according to the statistics, most internet users employ anti-viruses that are mostly run in the background and do not have any effect on their activities. Hence, along anti-viruses, anti-spyware can help to prevent potential spywares. File signatures must be updated regularly so that the above-mentioned approach can prevent attacks. Anti-spyware users should be aware that these kinds of software may also remove spywares that are the prerequisite of legal software performance. There are commercial software packs that can monitor the output traffic of the web- for a series of data definable by the user, which provide the users with desktop privacy services. In such situations, if the output traffic includes identification information of the user, the data are not be exported without user confirmation, even if the destination URL is not clear and it helps to prevent all social engineering attacks. The challenge confronting such an approach is to detect sensitive information and putting them in the Permit List and Deny List for specific websites.

## 6- DEALING WITH PHISHING BASED ON THE CERT MISSION

Electronic transactions and online banking are mainly designed for consumer convenience and reducing the cost of banking operations. But, a sharp increase in the electronic banking crimes has weakened the success of this plan and customers prefer physical presence to perform a secure transaction. Opponents of online transactions argue that phishing threats %86 of home users and %14 of home investors. At the global level, about 3000 phishing has been reported per month, more than %80 of which have to do with financial institutions [2]. The existence of security holes in the configuration of organizations' servers and not being aware of them, provides the context for phishing attacks. However, these types of attacks are designed so that reducing the sustained damages is almost impossible after their occurrence.

The CERT center is obliged to analyze different kinds of phishing and accordingly, present protective solutions based on the environmental variables of an organization, at the proper time and before attack occurrence. Coping with phishing can be based on three principles, based on CERT mission.

### 6-1. Informing

The main reason of phishing occurrence is the ignorance of internet users, and most approaches to cope with it require a training plan; hence, informing is regarded as the most important principle to fight against phishing. CERT center should include an educational campaign entitled "mechanisms for secure communications" in its organizational agendas, where it explained different types of phishing first and introduced all anti-phishing software and their method of updating as well as how to use software and hardware tokens to create disposable passwords. It should be noted that people who are in charge of providing electronic activities in banks and organizations are in priority as the target of these training courses, because training should be done in such a way that that security approaches be spread in a network manner by these individuals, and even naïve users also stay safe against such attacks.

In addition, CERT centers across the country should coordinate and create a common database to recognize and list fake websites and share it among the internet users so that professional users help to complement this database in addition to being aware of such addresses.

### 6-2. Support

After identifying the kind of phishing, the CERT center should analyze its reasons for an occurrence such as the unfamiliarity of internet users with strong technologies and algorithm of encryption. Based on the studies conducted in CERT center, nearly %80 of lay users cannot distinguish a phishing email from a safe and correct email. A kind of phishing attack which is regarded so destructive and insensible is that a third party hears them without the knowledge of sender or receiver during information transfer.

By using PGP (Pretty Good Privacy) software to digitally sign your messages, we can make sure that only the intended receiver will receive your message. PGP works based on encrypting the public key and has a pair of keys. To receive a private email from an individual, you should first provide that individual with the public key- which is an easily transferable block of text you have created using this software – and keep the private key for yourself. Know the individual encrypts his message using your public key, signs it with his private key and then send it to yourself. Now, after receiving email and using the sender's public key which you have been already provided with, you can make sure that it is legitimate and no change has been made on its text and decrypt it. It is obvious that if you are the sender of the email, all these processes will happen in reverse. Another capability of PGP software is to encrypt and decrypt the information of clipboard memory by which you can ensure the accuracy of email by copying and pasting links attached to.

S/MIME (Secure/ Multipurpose Internet Mail Extensions) is a universal standard which is used to encrypt emails. Using this service requires a third party license for both public uses over the internet and personal uses in private organizations. When you send email using this service, your email is, in fact, signed digitally and when it is received, the signature is reconciled with sender's information to ensure the identity of the sender. This service also allows for an optional encryption using public keys which are a synthesis of your digital license and password.

### 6-3. Computer help

Conducting a survey and collecting all causes of occurring a phishing attack, the CERT center should try to remove similar security weaknesses to prevent the occurrence of phishing via conventional methods. Because they are completely new, testing anti-phishing software by internet users has a high rate of risk, and some of these kinds of software detect phishing attack after their occurrence and when it is too late. The last version of popular browsers of Mozilla and Internet Explorer include a proper anti-phishing capability. Phishtank (phishtank.com) argues that it can detect phishing websites and offer solutions to fight against them, where users list phishing websites themselves and add them to their databases. The CERT center should test all available anti-phishing software in a comprehensive plan, and introduce those having higher success rate during the test. Some of these kinds of toolbar anti-phishing software are as follows: Windows Live Toolbar is available in IE7, is installed on IE6, and is produced by Microsoft having blocked %51 of phishing websites so far.

- Google Toolbar is a Google product, is available on Firefox, and has blocked %85 of phishing websites by now. This toolbar which is linked to Google database is updated every 2 hours.

- NetCraft Toolbar is produced by the English company of NetCraft which can be installed on Firefox and IE and has blocked %95 of phishing websites so far.

- NC Toolbar is produced by Symantec Company which can be installed on IE and has blocked %76 of fake websites so far. Preventing Key-logger software is of the advantages of this tool.

- eBay Toolbar is an eBay company product, specialized for using this website, enjoys a proper graphical interface. This tool has managed to block %90 of phishing attacks on eBay websites.

### 7- PROPOSED SOLUTIONS FOR CERT CENTERS TO PREVENT PHISHING ATTACK

Creating a common internet page for banks that contain links to all electronic payment portals is a proper solution to decrease theft through fishing. Based on this plan, a page entitled "Central Bank" is created with a special standard and logo. By entering this page, users can enter the portal of other banks. Creating just a page and logo of banks increases public confidence and in turn, can decrease phishing related crimes to a considerable extent. Now, when we search the name of a bank several portals and pages related to the respected bank open, the existence of multiple pages allows criminals to create phishing pages. It is emphasized that it is always easier to apply all the mentioned security approaches and protective layers on an internet page than to control the security of multiple bank portals over the internet.

### 8- CONCLUSION

Phishing is different from the first-grade traditional internet fraud, on a scale of deception that can really be committed to sabotaging. This kind of attack via email and World Wide Web that allows fraudulent to obtain thousands and millions of victims per minute without any cost was discussed in the present study.

The important principle in phishing is that the invader must continue to gain the trust of the victim for a successful attack.

Since there is no face-to-face communication between the invader and the victim, the clients have little information to decide to work with a legitimate email or website. The technical and ultimate solution to cope with phishing attacks is to make a significant change in internet infrastructure which is beyond the ability of any of the institutions to implement. In the present study, some approaches were proposed by which CERT centers can somehow cover the inability to change the infrastructure to create a protective layer against these attacks, and it is hoped that they are put into practice.

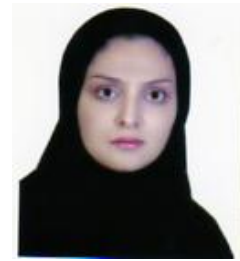

She received her B.E.  and M.S. degree in computer and Information Technology engineering from Yazd University, Iran in 2008 and 2010 respectively and is currently pursuing her Ph.D. in computer engineering in Sharif University of Technology, Iran. In 2011, she joined the department of computer and information technology engineering, Jahrom University, as an instructor. My research interests are computer network, its security and social networking in general.